\documentclass[pra, a4paper, twocolumn, showpacs,superscriptaddress]{revtex4-1}
\usepackage{amsmath}
\usepackage{amsfonts}
\usepackage{graphicx}
\usepackage{bbm}
\usepackage{longtable}
\begin{document}
\title{$N$ term pairwise correlation inequalities, steering and joint measurability}  
\author{H. S. Karthik}
\affiliation{Raman Research Institute, Bangalore 560 080, India} 
\author{A. R. Usha Devi}\email{arutth@rediffmail.com} 
\affiliation{Department of Physics, Bangalore University, 
Bangalore-560 056, India}
\affiliation{Inspire Institute Inc., Alexandria, Virginia, 22303, USA.}
\author{J. Prabhu Tej}
\affiliation{Department of Physics, Bangalore University, 
Bangalore-560 056, India}
\author{A. K. Rajagopal}
\affiliation{Inspire Institute Inc., Alexandria, Virginia, 22303, USA.}
\affiliation{Institute of Mathematical Sciences, C.I.T. Campus, Taramani, Chennai, 600113, India}
\affiliation{Harish-Chandra Research Institute, Chhatnag Road, Jhunsi, Allahabad 211 019, India.}
\author{Sudha}
\affiliation{Department of Physics, Kuvempu University, Shankaraghatta, Shimoga-577 451, India }
\affiliation{Inspire Institute Inc., Alexandria, Virginia, 22303, USA.}
\author{A. Narayanan}
\affiliation{Raman Research Institute, Bangalore 560 080, India}

\date{\today}
\begin{abstract} 
Chained inequalities involving  pairwise correlations of qubit observables in the equatorial plane are constructed based on the positivity of a sequence of moment matrices.  When  a jointly measurable  set of {\em fuzzy}  POVMs  is employed in first measurement of every pair of sequential  measurements, the chained pairwise correlations do not violate the {\em classical bound} imposed by the moment matrix positivity. We identify  that    {\em incompatibility} of the set of POVMs employed in first measurements is  only necessary, but not sufficient, in general, for the violation of the inequality. On the other hand, there exists a  one-to-one equivalence between the {\em degree of incompatibility} (which quantifies the joint measurability) of the equatorial qubit POVMs  and the optimal violation of a non-local steering inequality, proposed by Jones and Wiseman (Phys. Rev. A, {\bf 84}, 012110 (2011)).  To this end, we construct a {\em local} analogue of this steering inequality in a single qubit system and show that its violation is a  mere reflection of measurement incompatibility of equatorial qubit POVMs, employed in first measurements in the sequential unsharp-sharp scheme.      

\end{abstract}
\maketitle
\section{Introduction}
Conceptual foundations of quantum theory deviate drastically from the classical world view. The prominent counter intuitive features pointing towards the quantum-classical divide are a subject of  incessant  debate ever since the birth of quantum theory. Pioneering works by  Bell~\cite{bell}, Kochen-Specker~\cite{ks}, Leggett-Garg~\cite{lg} are significant in  bringing  forth the perplexing features arising within the quantum scenario, in terms of correlation inequalities, constrained to obey  {\em classical} bounds. Violation of the inequalities sheds light on the {\em the non-existence of a joint probability distribution} for the measurement outcomes of all the associated observables~\cite{fine, aru, sing}. 

In fact, non-commutativity of the observables forbids assignment 
of  {\em joint sharp realities} to their outcomes in  projective valued (PV) measurements. Subsequently, it is not possible to envisage a {\em bona fide} joint probability distribution for the outcomes of PV measurements of non-commuting observables. However,  the  generalized  measurement framework~\cite{Busch} goes beyond the conventional PV measurement scenario, where positive operator valued measures (POVMs) are employed.  Joint measurability (or {\em compatibility}) of a set of  POVMs is possible even when they do not commute. For declaring that a set of POVMs are jointly measurable there should exist  a {\em global} POVM,  measurement statistics of which enables one to retrieve that of the set of compatible  POVMs.  Within the purview of  generalized measurements, it is  possible to assign  {\em fuzzy joint realities} (and in turn, a valid joint probability distribution) to the statistical outcomes of non-commuting observables, when the corresponding POVMs are {\em compatible}. 

In recent years there is a surge of research activity dedicated to explore the notion of measurement incompatibility and its connection with the counter-intuitive quantum notions like non-locality, contextuality and non-macrorealism~\cite{Barnett05,Son05,Stano08,Wolf09,Yu10,LSW11,Reeb13,Kunjwal14,Brunner14,Guhne14, Guhne15, HSKJOSAB15,Pusey15,Heinosari15,HSKCurrSc,Chen16}. In particular, it is known that measurement incompatibility plays a key role in bringing to surface the violations of  the so-called no-go theorems in the quantum world. 
Wolf {\em et al.}~\cite{Wolf09} proved that a set of two incompatible dichotomic POVMs are necessary and sufficient to violate the Clauser-Horne-Shimony-Holt (CHSH) Bell inequality~\cite{CHSH}. However, this result may   not hold, in general, for Bell non-locality tests where more than two incompatible  POVMs with any number of outcomes are employed i.e., it is possible to identify a set of non-jointly measurable POVMs, which fail to reveal Bell-type non-locality, in general~\cite{Brunner16}. Interestingly, there exists a {\em one-to-one} equivalence~\cite{Brunner14,Guhne14,Guhne15} between measurement incompatibility and quantum steerability (i.e., Alice's ability to non-locally alter Bob's states  by performing local measurements on her part of the quantum state~\cite{Cavalcanti}). More specifically, a set of fuzzy  POVMs is said to be incompatible {\em if and only if} it can be used to show steering in a quantum state.

From the point of view of  an  entirely different mathematical perspective, the {\em classical moment problem}~\cite{tomarkin,akheizer,
hsk13,HSKCurrSc,HSK_AQIS13,HSK_unpublished}  addresses the issue of the existence of  probability distribution corresponding to a given a sequence of statistical moments. Essentially, {\em the classical moment problem} points out that a given sequence of real numbers qualify to be the  moment sequence of a {\em legitimate} probability distribution {\em if and only if} the corresponding moment matrix is positive.  In other words, {\em existence of a valid joint probability distribution},  consistent with the given  sequence of {\em moments}, necessitates positivity of the associated moment matrix. Moment matrix constructed in terms of pairwise correlations of observables in the quantum scenario is not necessarily positive~\cite{HSK_unpublished,HSK_AQIS13,HSKCurrSc} and thus, one witnesses violation of  Bell, Leggett-Garg, non-contextual inequalities (which can be realized to be the positivity constraints on the eigenvalues of the moment matrix).  In turn,  violation of these inequalities points towards  non-existence of joint probabilities corresponding to the measurements of all the observables employed. Moments extracted from  measurements of a set of POVMs result in a positive moment matrix, if the {\em degree of incompatibility} is restricted to lie within the range specified by the compatibility of the set of 
POVMs employed~\cite{HSKCurrSc}.             

In this paper, we construct  $N$ term chained  correlation  inequalities involving pairwise correlations of $N$ dichotomic random variables based on the positivity of a sequence of $4\times 4$ moment matrices. The  bound on the linear combination of pairwise  correlations (recognized through the positivity of moment matrices) ensures the existence of  joint probabilities for the statistical outcomes. When the dichotomic classical random variables are replaced by  qubit observables, one  witnesses a violation of the chained correlation inequalities~\cite{Wehner,Guhne13}. The maximum violation of the inequalities in the quantum scenario (i.e., the corresponding Tsirelson-like bound)  has been established in Refs.~\cite{Wehner,Guhne13}. The dichotomic observables, which result in the  maximum quantum violation, correspond to the qubit observables in a plane. Here, we investigate the {\em degree of incompatibility} necessary for the joint measurability of the equatorial plane noisy qubit POVMs (i.e., a mixture of qubit observable in the equatorial plane and the identity matrix). Based on this, we identify that the chained inequalities are always satisfied, when the equatorial noisy qubit POVMs, employed in first measurements of sequential pairwise measurements are all jointly measurable. But, incompatible POVMs are, in general, not sufficient for  violation of the chained inequalities for $N> 3$. On the other hand, we show that there is a one-to-one correspondence between the joint measurability of a set of  equatorial plane noisy qubit POVMs and the optimal violation of a linear non-local steering inequality proposed by Jones and Wiseman~\cite{Jones}. This leads us towards the construction of a {\em local} analogue of this steering inequality in a single qubit system --  violation of which gives an evidence for the non-joint measurability of the set of equatorial plane qubit POVMs, employed in first of every sequential pair measurements.  

We organize the contents of the  paper as follows: In Sec.~II we outline the notion of  compatible POVM. As a specific case, we discuss the compatibility of  qubit POVMs in the equatorial plane of the Bloch sphere and obtain the necessary condition for the unsharpness parameter quantifying the {\em degree of incompatibility}. Sec.~III is devoted towards  (i) formulation of chained $N$ term pairwise correlation inequalities constructed from the positivity of moment matrices;  (ii)  optimal violation of the inequalities  in the quantum scenario, when  qubit observables in the equatorial plane are employed and the connection between the degree of incompatibility  of POVMs, used in first of every sequential pair measurements, and the strength of violation of the correlation inequalities. In Sec.IV we show that  the  steering inequality proposed by Jones and Wiseman~\cite{Jones}  has a one-to-one correspondence with the joint measurability of equatorial qubit POVMs. A local analogue of this steering inequality for a single qubit system, involving  $N$ settings of sequential unsharp-sharp pairwise correlations is constructed. Sec.V contains a    summary of our results and concluding remarks.

\section{Joint measurability of POVMs}    

In the conventional quantum framework, measurements are described in terms of the spectral projection operators of the corresonding self-adjoint observables. And joint measurability of two commuting observables is ensured because results of a single PV measurement are comprised of those of both the observables. However, non-commuting observables are declared as incompatible under the regime of  PV measurements. Introduction of  POVMs in 1960's by Ludwig~\cite{Ludwig64}  and  subsequent investigations on their applicability~\cite{BuschLahti84}, led to a mathematically rigorous generalization of measurement theory. It is the notion of  compatibility (the notion of compatibility of a set of POVMs  will be defined in the following) -- rather than commutativity -- which gains importance so as to recognize if a given set of POVMs are jointly measurable or not~\cite{note1}.
   
A POVM  is a set $\mathbbm{E}_x=\{E_x(a)=M^\dag_x(a)\, M_x(a)\}$ comprising of positive self-adjoint operators 
$0\leq E_x(a)\leq \mathbbm{1}$,  satisfying  
$\sum_a E_x(a)=\sum_a\, M^\dag_x(a) M_x(a)=\mathbbm{1}$;  $a$ denotes the outcome of measurement  and $\mathbbm{1}$ is the identity operator.    Under measurement $\{M_x(a)\}$ a quantum system, prepared in the state $\rho$, undergoes a positive trace-preserving generalized L$\ddot{\rm u}$der's transformation i.e., 
\begin{equation}
\rho\longmapsto \sum_a\, M_x(a)\, \rho\, M^\dag_x(a),  
\end{equation}
and an outcome  $a$  occurs with probability $p(a\vert x)={\rm Tr}[\rho\, M^\dag_x(a)M_x(a)]={\rm Tr}[\rho\, E_x(a)]$. 
Results of PV measurements can be retrieved as a special case, when the the POVM $\{E_x(a)\}$ consists of complete, orthogonal projectors.  

A finite collection $\{\mathbbm{E}_{x_1}, 
\mathbbm{E}_{x_2}, \ldots, \mathbbm{E}_{x_N}\}$ of $N$ POVMs is said to be  jointly measurable (or compatible), if there exists a {\em grand} POVM $\mathbbm{G}=\{G(\lambda);\ 0\leq G(\lambda)\leq \mathbbm{1},\, \sum_{\lambda}\, G(\lambda)=\mathbbm{1}\}$, with  outcomes denoted by a collective index $\lambda\equiv 
\{a_1,a_2,\ldots, a_N\}$, such that the individual POVMs $\mathbbm{E}_{x_i}$ can be expressed as its marginals~\cite{Stano08}:
\begin{eqnarray} 
\label{pp} 
E_{x_k}(a_k)&=&\sum_{\begin{array}{ll} a_1,a_2,\ldots, a_{k-1},\\ 
 a_{k+1},\ldots, a_N\end{array}}\,   G(\lambda=\{a_1,a_2,\ldots, a_N\}), \nonumber \\   
\end{eqnarray}
for all $k=1,2,\ldots N$.
From now on, we denote the collective index $\lambda=\{a_1,a_2\ldots, a_N\}$ characterizing measurement outcomes of global POVM $\mathbbm{G}$ by  ${\bf a}=(a_1,a_2,\ldots, a_N)$ for brevity.  

When a measurement of the global POVM $\mathbbm{G}\equiv\{G({\bf a})\}$ is carried out in an arbitrary quantum state $\rho$, an outcome `${\bf a}$' occurs  with probability ${\rm Tr}[\rho\, G({\bf a})]=p({\bf a})$ . Then, the corresponding results  $(p(a_k\vert x_k), a_k)$  (viz., the outcomes $a_k$ and the probabilities $p(a_k\vert x_k)={\rm Tr}[\rho\, E_{x_k}(a_k)]$),  for all the compatible POVMs $\mathbbm{E}_{x_k}$ can be deduced by post-processing the collective measurement data~\cite{note1} 
$(p({\bf a}),\{{\bf a}\})$ of the global POVM  $\mathbbm{G}$:
\begin{eqnarray} 
\label{prob} 
p(a_k\vert x_k)&=&\sum_{\begin{array}{ll} a_1,a_2,\ldots, a_{k-1},\\ 
 a_{k+1},\ldots, a_N\end{array}}\, p({\bf a}).   
\end{eqnarray}
A set of POVMs $\{\mathbbm{E}_{x_k}\},\ k=1,2,\ldots , N$ are declared to be compatible iff they are {\em marginals} of a global POVM $\mathbbm{G}$ (as expressed in (\ref{pp})).    
            
\subsection{Example of noisy qubit POVMs:} Consider a pair of  qubit observables   
$\sigma_x=\displaystyle\sum_{a_x=\pm 1}\, a_x\, \Pi_{x}(a_x)$ and $\sigma_z=\displaystyle\sum_{a_z=\pm 1}\, a_z\, \Pi_{z}(a_z)$. Sharp PV measurements of these self adjoint observables $\sigma_x$, $\sigma_z$ are incorporated in terms of their spectral projectors, 
\begin{eqnarray}
\label{sharpxy}
 \Pi_{x}(a_x)&=& \frac{1}{2}\left(\mathbbm{1}+ a_x\, \sigma_x \right),\nonumber \\
 \Pi_{z}(a_z)&=& \frac{1}{2}\left(\mathbbm{1}+  a_z\, \sigma_z\right). 
 \end{eqnarray}
Within the conventional framework of  PV measurements, the non-commuting qubit observables $\sigma_x$ and $\sigma_z$ are not jointly measurable. However, it is possible to consider a particular choice of jointly measurable noisy qubit POVMs $\mathbbm{E}_{x}=\{E_{x}(a_x)\}, \mathbbm{E}_{z}=\{E_{x}(a_x)\}$, by mixing white noise to the respective projection operators i.e., 
\begin{eqnarray}
\label{unsharpxy}
 E_{x}(a_x)&=&  \eta\ \Pi_{x}(a_x) + (1-\eta)\, \frac{\mathbbm{1}}{2} \nonumber \\ 
& =&\frac{1}{2}\left(\mathbbm{1}+ \eta\, a_x\, \sigma_x \right) \nonumber \\
E_{z}(a_z)&=&  \eta\ \Pi_{z}(a_z) +(1-\eta)\, \frac{\mathbbm{1}}{2} \nonumber \\ 
&=&\frac{1}{2}\left(\mathbbm{1}+ \eta\, a_z\, \sigma_z \right). 
 \end{eqnarray} 
where $0\leq \eta\leq 1$ denotes the unsharpness parameter. When  $\eta=1$, the noisy qubit POVMs (\ref{unsharpxy}) reduce to their corresponding {\em sharp} PV counterparts. Throughout this paper, we will be focusing on the joint measurability (compatibility) of noisy qubit observables of the form given by (\ref{unsharpxy}).      

The dichotomic POVMs  $\mathbbm{E}_{x},\ \mathbbm{E}_{z}$ are jointly measurable if there exists a {\em four outcome} global POVM 
$\mathbbm{G}=\{G(a_x,a_z);\   a_x=\pm 1,\,a_z=\pm 1\}$,  such that     
\begin{eqnarray}
\label{pcond}
\sum_{a_z=\pm 1}\,  G(a_x,a_z)&=&E_{x}(a_x)\nonumber \\ 
\sum_{a_x=\pm 1}\,  G(a_x,a_z)&=&
E_{z}(a_z) \nonumber \\ 
\sum_{a_x,a_z=\pm 1}\, G(a_x,a_z)&=&\mathbbm{1}, \ \
G(a_x,a_z)\geq 0.  
\end{eqnarray} 
It has been shown~\cite{Busch, Stano08} that the POVMs $\mathbbm{E}_{x},\ \mathbbm{E}_{z}$ are jointly measurable in the range $0\leq \eta\leq 1/\sqrt{2}$ i.e., it is possible to construct a global POVM $\mathbbm{G}$ comprised of the elements  
\begin{equation}
G(a_x,a_z)=\frac{1}{4}\left(\mathbbm{1}+\eta\, a_x\, \sigma_x + \eta\, a_z\, \sigma_z\right),\  0\leq \eta\leq 1/\sqrt{2}
\end{equation}
which obey (\ref{pcond}).  

Similarly, triple-wise joint  measurements of the qubit observables $\sigma_x,\ \sigma_y$ and $\sigma_z$ 
could be  envisaged by considering  the fuzzy  POVMs 
$\mathbbm{E}_{x}, \mathbbm{E}_{y},$  $\mathbbm{E}_{z}$, elements of which are given respectively by,    
\begin{eqnarray*}
E_{x}(a_x)&=&\frac{1}{2}\left(\mathbbm{1}+ \eta\, a_x\, \sigma_x \right);\, a_x=\pm 1, \\ 
E_{y}(a_y)&=&\frac{1}{2}\left(\mathbbm{1}+ \eta\, a_y\, \sigma_y \right);\, a_y=\pm 1, \\
E_{\sigma_z}(z)&=&\frac{1}{2}
\left(\mathbbm{1}+ \eta\, a_z\, {\sigma_z} \right);\, a_z=\pm 1 
\end{eqnarray*}
 in the range $0\leq \eta\leq 1/\sqrt{3}$ of the unsharpness parameter~\cite{Stano08, LSW11}. 

In general, the necessary condition  on the 
{\em unsharpness parameter} such that the qubit POVMs 
$\{E_{x_k}(a_k=\pm 1)=\frac{1}{2}\ [\mathbbm{1}+\eta\, a_k\, \vec{\sigma}\cdot\hat{n}_k],\ \ k=1,2,\ldots N\}$ are jointly measurable is derived in Ref.~\cite{LSW11,Ravi}:  
\begin{equation} 
\label{etaopt}
\eta\leq \frac{1}{N}\, \stackrel{\rm \ \ \ \large max\ \ \ \ } {_{{\bf a}}}\,\, \vert \vec{m}_{{\bf a}}\vert,
\end{equation}
where $\vec{m}_{{\bf a}}$ is defined by,   
\begin{equation}
\label{mdef}
\vec{m}_{{\bf a}}=\sum_{k=1}^{N}\,  \hat{n}_k\, a_k, \ a_k=\pm 1.
\end{equation} 
The maximization is carried out over all the  $2^N$ outcomes~\cite{note2}\  $\{{\bf a}=(a_1=\pm 1,a_2=\pm 1,\ldots, a_N=\pm 1)\}$. A sufficient condition places the following constraint on the unsharpness parameter (derived in Ref.~\cite{LSW11}):  
\begin{equation}
\label{etasuff}
\eta\leq \frac{2^N}{\sum_{\bf a}\,\vert\vec{m}_{\bf a}\vert}. 
\end{equation}
 
In Table.~I we list the optimal value $\eta_{\rm opt}$  of the  unsharpness parameter ( evaluated using (\ref{etaopt}), (\ref{etasuff})) --  below which the  qubit POVMs $\{E_{x_k}(a_k)=\frac{1}{2}(I+\eta\, a_k\, \vec{\sigma}\cdot \hat{n}_k); a_k=\pm 1\}$ are jointly measurable  -- in the specific cases~\cite{note3} of  $\hat{n}_k, k=1,2,3$ constituting (i) orthogonal axes and (ii) trine axes~\cite{LSW11, Ravi} i.e., three coplanar unit vectors with  $\hat{n}_k\cdot \hat{n}_l=-\frac{1}{2},\ k\neq l=1,2,3$. 
\begin{table} \label{tab1}
\begin{tabular}{|c |c |c|}
\hline 
Number  of & Orientation    & $\eta_{\rm opt}$  \\
 POVMs    & of $\hat{n}_k$  &   \\ 
\hline 
   &  Orthogonal axes  & \\ 
& & \\
$N=3$      &  $\hat{n}_k\cdot \hat{n}_l=0,\  k\neq l=1,2,3$ 
			&$\frac{1}{\sqrt{3}}$ \\ 	
			& & \\
$N=2$ &      $\hat{n}_1\cdot \hat{n}_{2}=0$ &   
$\frac{1}{\sqrt{2}}$ \\
& & \\ 
\hline
                		   &    Trine axes & \\   
												& & \\
$N=3$  & $\hat{n}_k\cdot \hat{n}_{l}=-\frac{1}{2};\, k\neq l=1,2,3$ 
&  $\frac{2}{3}$  \\ 
& & \\
$N=2$ &      $\hat{n}_1\cdot \hat{n}_{2}=-\frac{1}{2}$  &   
$0.732$ \\ 
& & \\
\hline
\end{tabular}
\caption{Optimal value $\eta_{\rm opt}$ of  the unsharpness parameter (evaluated using  the necessary and sufficient conditions (\ref{etaopt}), (\ref{etasuff})), below which the joint measurability of the qubit POVMs 
$\{E_{x_k}(a_k)~=~\frac{1}{2}(\mathbbm{1}+\eta\, a_k\,\vec{\sigma}\cdot\hat{n}_k)\}$ for  different orientations $\hat{n}_k$ are compatible.}
\end{table}

\subsection{Joint measurability of equatorial qubit observables}

Here we consider joint measurability of $N$ equatorial qubit observables $\sigma_{\theta_k}=\sigma_{x}\, \cos(\theta_k)+\sigma_{y}\, \sin(\theta_k),$ $\theta_k=k\, \pi/N,\  k=1,2,\ldots, N$, which  correspond geometrically to the points on the circumference of the  circle  in the equatorial half-plane ($z=0$) of the Bloch sphere, separated  successively by an angle $\theta=\pi/N$. 

Consider equatorial qubit POVMs $\mathbbm{E}_{\theta_k}$, elements of which are given by,  
\begin{eqnarray}
\label{ethetak}
 E_{\theta_k}(a_k=\pm 1)&=& M^\dag_{\theta_k}(a_k)M_{\theta_k}(a_k) \nonumber \\
&=&\frac{1}{2}(\mathbbm{1}+\eta\, a_k\, \sigma_{\theta_k}).
\end{eqnarray}   
When the set $\{\mathbbm{E}_{\theta_k}\}$ of POVMs is compatible, there exists a global qubit POVM
$\mathbbm{G}$  comprised of $2^N$ elements 
\begin{eqnarray*}
G(\mathbf{a})=\frac{1}{2^N}\left(\mathbbm{1}+\eta\, \sum_{k=1}^N\,  a_k\, \sigma_{\theta_k}\right),
\end{eqnarray*}
with  $0\leq \eta\leq \eta_{\rm opt}$.   Using (\ref{etaopt}), (\ref{etasuff}), we have evaluated the range of unsharpness parameter $0\leq \eta\leq \eta_{\rm opt}$, such that the 
POVMs $\mathbbm{E}_{\theta_k}, k=1,2,\cdots N$ are jointly measurable. Based on our computations of   $\eta_{\rm opt}$,  for  small values of $N$, we recognize the  following cut-off  $\eta\leq \eta_{\rm opt}$  for any $N$: 
\begin{equation}
\label{etaeq}
\eta_{\rm opt}=\frac{1}{N}\sqrt{N+2\, 
\sum_{k=1}^{[\frac{N}{2}]}\, (N-2k)\, \cos\left(\frac{k\,\pi}{N} \right)}.
\end{equation} 
The values of $\eta_{\rm opt}$ are listed in Table~II. 
In the large $N$ limit, the degree of incompatibility (i.e., the cut-off value of the unsharpness parameter)  approaches $\eta_{\rm opt}\rightarrow 0.6366$ and thus the POVMs associated with the set of all  qubit observables $\sigma_\theta$, $0\leq \theta\leq \pi$  in the equatorial plane of the Bloch sphere are  jointly measurable in the range $0\leq \eta^{(\infty)}_{\rm opt}\leq 0.6366.$    

More recently~\cite{Uola} Uola {\em et.al.} investigated incompatibility of some noisy observables in finite dimensional Hilbert spaces by developing a new technique -- which they referred to as {\em adaptive strategy}. In particular, they independently identified the following sufficient condition for the simultaneous measurements of  qubit observables in a plane $\sigma_{\theta_k}=\sigma_{x}\, \cos(\theta_k)+\sigma_{y}\, \sin(\theta_k),$ $\theta_k=k\, \pi/N,\  k=1,2,\ldots, N$ based on their approach~\cite{Uola}: 
\begin{eqnarray} 
\label{Uola} 
\eta &\leq& \frac{2}{N}\, \sum_{k=1}^{[\frac{N}{2}]}\, \cos\left(\frac{(2k-1)\,\pi}{2N}\right)\nonumber \\ 
&=&\frac{1}{N\, \sin(\pi/2N)},
\end{eqnarray}
which too agrees perfectly with the optimal value (\ref{etaeq}) of the unsharpness parameter (see Table~II).    
        
\begin{table} \label{tab2}
\begin{tabular}{|c |c|}
\hline 
& \\
\hskip 0.2in Number  of \hskip 0.2in & $\eta_{\rm opt}$    \\
 POVMs       &   \\
& \\ 
\hline 
 3      & \ \  0.6666 \ \ \\ 
4 & \ \   0.6532 \ \ \\ 
5&  \ \  0.6472 \ \ \\ 
6 &  \ \  0.6439 \ \ \\ 
10 & \ \ 0.6392 \ \ \\ 
20 & \ \ 0.6372 \ \ \\
50 & \ \  0.6367 \ \ \\ 
100 & \ \ 0.6366 \ \ \\ 
\hline 
\end{tabular}
\caption{Optimal value   $\eta_{\rm opt}$ of the unsharpness parameter (see (\ref{etaeq})) specifying the 
    joint measurability of the  equatorial qubit observables
		 $\sigma_{\theta_k}=\sigma_x\, \cos(k\, \pi/N)+\sigma_{y}\,\sin(k\, \pi/N);$  \ $k=1,2,\ldots, N.$}
\end{table}

\section{Chained $N$ term correlation inequalities and joint measurability}

The local realistic framework places bounds on correlations between the outcomes of measurements, carried out by spatially separated parties  and   Bell inequalities formulated in terms of these correlations get violated in the framework of quantum theory. On the other hand, quantum theory too places a strict limit  on the strength of these correlations. The maximum violation of  CHSH inequality~\cite{CHSH}, by non-local quantum correlations, is constrained by the Tsirelson bound~\cite{Tsirelson} $2\sqrt{2}$. The CHSH inequality involves measurements of two pairs of dichotomic observables on a bipartite system (denoted by $(A_1,\, A_2)$ and $(B_1,\, B_2)$ which are local observables measured by  Alice, Bob respectively) and four correlation terms:
\begin{equation}
\label{chsh}
\langle A_1\, B_1\rangle+\langle A_1\, B_2\rangle+\langle A_2\, B_1\rangle-\langle A_2\, B_2\rangle\leq 2
\end{equation} 
An interesting connection between joint measurability and violation of the CHSH inequalities, within the framework of quantum theory, has been brought out recently by Banik {et. al.}~\cite{Banik}.  In general, they showed that, in a no-signaling probabilistic theory the maximum strength of violation of the inequality (\ref{chsh}) by any  pair of  $(A^\eta_1,\, A^\eta_2)$ of quantum dichotomic observables (unsharp counterparts of  $(A_1,\, A_2)$) is essentially determined by the optimal degree of incompatibility $\eta_{\rm opt}$ which, in quantum theory, is identified to be $\frac{1}{\sqrt{2}}$  i.e.,    
\begin{eqnarray}
\label{chsh1}
\langle A^\eta_1\, B_1\rangle+\langle A^\eta_1\, B_2\rangle+\langle A^\eta_2\, B_1\rangle-\langle A^\eta_2\, B_2\rangle&\leq 
&\frac{2}{\eta_{\rm opt}} \nonumber \\
    &=& 2\sqrt{2}.  \nonumber \\
\end{eqnarray} 
In other words, the degree of incompatibility $\eta_{\rm opt}=\frac{1}{\sqrt{2}}$ of measurements in the quantum framework is shown to place limitations on the maximum strength of violations of the four term CHSH inequality, by retrieving the quantum Tsirelson bound 
$2\sqrt{2}$.       

Does this connection between the degree of incompatibility and the  Tsirelson-like bound (maximum strength of violation),  hold in general, when more than two incompatible  measurements are involved? We explore this question through  $N$ term correlation inequalities, which we  formulate from positivity of a sequence of moment matrices.

Consider $N$ classical random variables $X_k,\, k=1,2,\ldots N$ with  outcomes $a_k=\pm 1$. Let $\xi^T_{k}=\left( 1,\ 
 a_1a_{k},\  a_ka_{k+1}, \ a_1 a_{k+1}\right),  \ k=2,3,\ldots N-1$ denote row vectors.   
We construct a sequence of $4\times 4$ moment matrices  $M_{k}=\langle \xi_{k}\,\xi^T_{k}\rangle$,  expressed explicitly as, 
{\footnotesize\begin{eqnarray} 
M_{k}=\left(\begin{array}{cccc}
1 & \langle X_1\, X_{k}\rangle & \langle X_{k}\, X_{k+1}\rangle & \langle X_{1}\, X_{k+1}\rangle \\ 
 \langle X_1\, X_{k}\rangle & 1 & \langle X_{1}\, X_{k+1}\rangle & \langle X_{k}\, X_{k+1}\rangle \\  
\langle X_{k}\, X_{k+1}\rangle & \langle X_{1}\, X_{k+1}\rangle & 1 &  \langle X_{1}\, X_{k}\rangle \\ 
\langle X_{1}\, X_{k+1}\rangle & \langle X_{k}\, X_{k+1}\rangle &   \langle X_{1}\, X_{k}\rangle & 1 \\ 
\end{array}\right), \nonumber \\ 
\end{eqnarray}} 
where $\langle X_k\, X_{l}\rangle, \ k\neq l$ denote  pairwise correlations of the variables $X_k$, $X_l$. (Here $\langle \cdot \rangle$ denotes the expectation value).  

In the classical probability setting, the moment matrix is, by construction,  real symmetric and positive semidefinite. 
The  eigenvalues $\lambda^{(k)}_i,\  i=1,2,3,4$ of the moment matrix are given by 
\begin{eqnarray}
\label{eig}
\lambda^{(k)}_1&=&1+\langle X_1\, X_{k}\rangle-\langle X_{k}\, X_{k+1}\rangle-\langle X_{1}\, X_{k+1}\rangle \nonumber \\ 
\lambda^{(k)}_2&=&1-\langle X_1\, X_{k}\rangle+\langle X_{k}\, X_{k+1}\rangle- \langle X_{1}\, X_{k+1}\rangle \nonumber \\ 
\lambda^{(k)}_3&=&1-\langle X_1\, X_{k}\rangle-\langle X_{k}\, X_{k+1}\rangle+\langle X_{1}\, X_{k+1}\rangle \nonumber \\
\lambda^{(k)}_4&=&1+\langle X_1\, X_{k}\rangle+\langle X_{k}\, X_{k+1}\rangle+\langle X_{1}\, X_{k+1}\rangle. \nonumber \\  
\end{eqnarray} 

Replacing  classical random variables $X_k$  by quantum dichotomic observables ${\mathbf X}_k=\vec{\sigma}\cdot\hat{n}_k, k=1,2,\ldots, N$ with eigenvalues $\pm 1$, and the classical probability distribution  by a density matrix, the moment matrix positivity results in linear constraints on  pairwise correlations of the observables  measured sequentially.    

Based on the positivity of a sequence of $N-1$  moment matrices $M_{2}, M_{3},\ldots, M_{N-1}$  one  obtains the inequalities  
$\displaystyle\sum_{k=2,3,\ldots, N-1} \lambda^{(k)}_i\geq 0,$ for the sum of the eigenvalues (see (\ref{eig})), which correspond to the following  chained inequalities involving pairwise correlations:
\begin{eqnarray}
\label{ch1} 
\sum_{k=2}^{N-1}\, &&\langle {\mathbf X}_k\, {\mathbf X}_{k+1}\rangle+ \langle {\mathbf X}_1\, {\mathbf X}_{N}\rangle  - \langle {\mathbf X}_1\, {\mathbf X}_{2}\rangle\leq N-2 \\
\label{ch2}
2\, \sum_{k=2}^{N}\, &&\left[\langle {\mathbf X}_1\, {\mathbf X}_{k}\rangle - \langle {\mathbf X}_k\, {\mathbf X}_{k+1}\rangle\right] +  \langle {\mathbf X}_{1}\, {\mathbf X}_{N}\rangle\leq N-2 \\
\label{ch3}
\sum_{k=1}^{N-1}\, &&\langle {\mathbf X}_{k}\, {\mathbf X}_{k+1}\rangle - \langle {\mathbf X}_{1}\, {\mathbf X}_{N}\rangle\leq N-2 \\
\label{ch4}
\sum_{k=2}^{N-1}\, &&\langle {\mathbf X}_{k}\, {\mathbf X}_{k+1}\rangle + 2 \sum_{k=2}^{N-2}\langle {\mathbf X}_{1}\, {\mathbf X}_{k+1}\rangle + \langle {\mathbf X}_{1}\, {\mathbf X}_{N}\rangle \leq N-2. \nonumber \\ 
\end{eqnarray}
Violation of these inequalities imply at least one of the moment matrix $M^{(k)}$ is not positive --  which in turn highlights {\em non-existence of a valid  joint probability distribution} for the outcomes of all the observables employed. However, it may be identified that by employing unsharp measurements of the  observables -- within their joint measurability region -- one can retrieve positivity of the sequence of moment matrices, and consequently,  the chained inequalities (\ref{ch1})-(\ref{ch4}) are satisfied. 

In particular, (\ref{ch3}) is analogous to the  $N$-term temporal correlation inequality investigated by Budroni {\em et. al.}~\cite{Guhne13}.  The  pairwise correlations $\langle {\mathbf X}_{k}\, {\mathbf X}_{k+l}\rangle$ arise from  the sequential measurements of the observables ${\mathbf X}_{k}$ and  ${\mathbf X}_{k+l}$ in a single quantum system.  Such inequalities involving sequential pairwise correlations of observables in a single quantum system (in contrast to correlations of the outcomes of local measurements at different ends of a spatially separated bipartite system as in (\ref{chsh})) have been well explored to highlight quantum contextuality~\cite{KCBS} and non-macrorealism~\cite{lg, aru,nori}.  
  
Budroni {\em et. al.}~\cite{Guhne13}  computed the maximal achievable value (Tsirelson-like bound) of  the left hand side of the 
chained $N$ term temporal correlation inequality (\ref{ch3}) and obtained 
\begin{equation}
\label{qtb}
{\cal S}^{Q}_{N}= \sum_{k=1}^{N-1}\, \langle {\mathbf X}_{k}\, {\mathbf X}_{k+1}\rangle_{\rm seq} - \langle {\mathbf X}_{1}\, {\mathbf X}_{N}\rangle_{\rm seq}
\leq N\, \cos\left(\frac{\pi}{N}\right). 
\end{equation} 
The classical bound $N-2$ on the chained $N$ term inequality (\ref{ch3}) can get violated in the quantum framework and a maximum value of $N\, \cos\left(\frac{\pi}{N}\right)$ could be achieved by choosing sequential measurements of appropriate observables. In particular, when a single qubit is prepared in a maximally mixed state $\rho=\mathbbm{1}/2$, sequential PV measurements of the observables 
$\sigma_{\theta_k}~=~\sigma_{x}\, \cos(\theta_k)+\sigma_{y}\, \sin(\theta_k),$ $\theta_k=k\, \pi/N,\  k=1,2,\ldots, N$  lead to pairwise correlations
\begin{eqnarray}
\label{pvalue} 
\langle {\mathbf X}_k\, {\mathbf X}_{k+l}\rangle_{\rm seq}&=&\langle \sigma_{\theta_k}\, \sigma_{\theta_{k+l}}\rangle_{\rm seq}\nonumber \\ 
&=&\cos(\theta_{k+l}-\theta_k) \nonumber \\ 
&=&\cos\left(\frac{l\, \pi}{N}\right). 
\end{eqnarray} 
Substituting (\ref{pvalue}) in (\ref{qtb}) we obtain the quantum Tsirelsen-like bound 
${\cal S}^{Q}_{N}=N  \cos\left(\frac{\pi}{N}\right).$

It is pertinent to point out that the observables $\{\sigma_{\theta_k}~=~\sigma_{x}\, \cos(\theta_k)+\sigma_{y}\, \sin(\theta_k),$ $\theta_k=k\, \pi/N,\  k=1,2,\ldots, N\}$ {\em need not}, in general, be associated with any particular time evolution; they are considered to be any ordered set of observables. Moreover, the pairs of sequential measurements are performed in independent statistical trials i.e., the
input state in every first measurement of the pair is $\rho=\mathbbm{1}/2$.
 
\subsection{Degree of incompatibility and violation of the chained correlation inequality (\ref{ch3})} 

It is seen that the average pairwise correlations $\langle {\mathbf X}_k\, {\mathbf X}_{k+l}\rangle_{\rm seq}$ of qubit observables 
${\mathbf X}_k\equiv\sigma_{\theta_k}, k=1,2,\cdots , N$, evaluated based on the results of sequential sharp PV measurements, lead to maximal violation of  the chained correlation inequality (\ref{ch3}). Instead of sharp PV measurements of the observables, we consider here an alternate sequential measurement scheme. We {\it separate} the set of observables $\{{\mathbf X}_k\equiv \sigma_{\theta_k},\ k=1,2,\cdots , N\}$ of first measurements of every sequential pair. We ask {\it if the chained inequality (\ref{ch3}) is violated, when measurement of first observables of every pair correlation  $\langle {\mathbf X}_k\, {\mathbf X}_{k+l}\rangle_{\rm seq}$ is done using noisy POVMs, while sharp PV measurements are employed for second observables in the sequence.}  Interestingly, we identify  that the chained inequality (\ref{ch3}) is {\em not violated}, whenever a compatible set of POVMs $\{\mathbbm{E}_{\theta_k}, k=1,2,\cdots, N\}$  (see \ref{ethetak}) is employed to carry out measurements of first observables of every sequential pair  -- irrespective of the fact that second measurements are all  sharp (and hence incompatible). In other words, incompatibility of the set of POVMs, employed in carrying out first measurements in the sequential scheme, is sufficient to witness violation of the chained inequality (\ref{ch3}). We now proceed to describe the sequential measurement scheme explicitly in the following.     
 
Consider $N$ noisy qubit observables $\mathbbm{E}_{\theta_k}$ with elements 
$\{E_{\theta_k}(a_k=\pm 1)=M^\dag_{\theta_k}(a_k)M_{\theta_k}(a_k)\}$ given by (\ref{ethetak}).  
From our discussions in Sec.~IIB,  it is seen that there exists a global qubit POVM 
$\mathbbm{G}$, when  $\eta$ lies in the range $0\leq \eta\leq \eta_{\rm opt}$ (see  
(\ref{etaeq}), for the values of the parameter $\eta_{\rm opt}$), such that the POVMs $\mathbbm{E}_{\theta_k}, k=1,2,\cdots , N$ are all jointly measurable.    

As before, we consider the initial state of the qubit to be $\rho=\mathbbm{1}/2$, a maximally mixed state. Carrying out unsharp measurement   $M_{\theta_k}(a_k)$, yielding  an outcome $a_k$, the initial state  gets transformed to 
\begin{eqnarray}
\label{stchange} 
\rho \rightarrow  \rho_{a_k}&=&\frac{M_{\theta_k}(a_k)\, \rho\, M^\dag_{\theta_k}(a_k)}{p(a_k\vert \theta_k)} \nonumber \\
            &=& \frac{1}{2}(\mathbbm{1}+\eta\, a_k\, \sigma_{\theta_k}), 
\end{eqnarray} 
where we have denoted 
${\rm Tr}[\rho\, M^\dag_{\theta_k}(a_k)M_{\theta_k}(a_k)]={\rm Tr}[\rho\, E_{\theta_k}(a_k)]=p(a_k\vert \theta_k)$. 
Following this with a second PV measurement of $\sigma_{\theta_{k+l}}$, on  the state $\rho_{a_k}$ results in the pairwise correlations 
\begin{eqnarray}
\label{etaineq}
\langle {\mathbf X}^{(\eta)}_k\, {\mathbf X}_{k+l}\rangle_{\rm seq}
&=&\sum_{a_k} p(a_k\vert \theta_k) \, {\rm Tr}[\rho_{a_k}\,  \sigma_{\theta_{k+l}}] \nonumber \\
&=& \eta\, \cos(\theta_{k+l}-\theta_k) \nonumber \\
&=&\eta\, \cos(\pi\, l/N). 
\end{eqnarray} 
So,  the left hand side of  chained  correlation 
inequality (\ref{ch3}) assumes the value,
\begin{eqnarray}
{\cal S}^{Q}_{N}(\eta)&=& \sum_{k=1}^{N-1}\, \langle {\mathbf X}^{(\eta)}_{k}\, {\mathbf X}_{k+1}\rangle_{\rm seq} - \langle {\mathbf X}^{(\eta)}_{1}\, {\mathbf X}_{N}\rangle_{\rm seq} \nonumber \\ 
 &=& \eta\,  N\, \cos\left(\frac{\pi}{N}\right),  
\end{eqnarray}  
when pairwise unsharp-sharp measurements of equatorial qubit observables are carried out.  
Within the joint measurability domain of the set  $\{\mathbbm{E}_{\theta_{k}}, k=1,2,\ldots N\}$ of first unsharp measurements in this sequential scheme, the sum of pairwise correlations obey 
\begin{eqnarray}
\label{achbound}
{\cal S}^{Q}_{N}(\eta)&=&\sum_{k=2}^{N-1}\, \langle X^{(\eta)}_{k}\, {\mathbf X}_{k+1}\rangle_{\rm seq} - \langle X^{(\eta)}_{1}\, {\mathbf X}_{N}\rangle_{\rm seq} 
\nonumber \\ && \ \ \leq 
\eta_{\rm opt}\,  N\, \cos\left(\frac{\pi}{N}\right).
\end{eqnarray}   
Using the optimal values $\eta_{\rm opt}$  specifying the degree of incompatibility of the equatorial qubit observables (see (\ref{etaeq}) and the values listed in Table~II), we evaluated  the maximum value
${\cal S}^{Q}_{N}(\eta_{\rm opt})$ $=\eta_{\rm opt}\, N\, \cos\left(\frac{\pi}{N}\right)$  attainable by the left hand side of the inequality (\ref{achbound}) for different values of $N$; these values are listed together with the corresponding classical and quantum bounds in Table~\ref{tab3}. It is evident that as the number of measurements $N$ increases, the quantum Tsirelson-like bound approaches the algebraic maximum value $N$, while the maximum achievable value of (\ref{achbound}) approaches ${\cal S}^{Q}_{N}(\eta_{\rm opt})\rightarrow 0.6366 \times N$ (which is equal to the classical bound $N-2=1$ for $N=3$ and is  less than
$N-2$ for $N>3$). More specifically, the classical bound is always satisfied, when the first measurements in the sequential scheme are carried out by compatible  POVMs. However, unlike the situation in the  CHSH-Bell inequality~\cite{Banik, Wolf09, notebell}  (\ref{chsh}), the maximum achievable value  
${\cal S}^{Q}_{N}(\eta_{\rm opt})$ is {\it not identically equal} to the classical bound of $N-2$, except in the case of $N=3$~\cite{HSKCurrSc}. So, it is evident that incompatible set  $\{\mathbbm{E}_{\theta_k};\  \eta > \eta_{\rm opt},\ k=1,2,\cdots N\}$ of POVMs  are necessary, but not sufficient to violate  the  chained $N$ term  correlation inequality (\ref{ch3}). Is it possible to find a steering protocol, for which  incompatibility of equatorial qubit measurements is both necessary and sufficient? In the next section we discuss a linear steering inequality involving equatorial  qubit observables~\cite{Jones} and  unravel how  violation of the inequality gets intertwined with measurement incompatibility.          

\begin{table} \label{tab3}
\begin{tabular}{|c|c|c|c|}
\hline 
&  &  & \\
\hskip 0.2in No.  of \ \ \hskip 0.2in & Classical  &  Quantum  & Maximum   \\ 
 POVMs  & bound  & bound   & achievable value 
 \\ 
employed & N-2 &  $N\, \cos\left(\frac{\pi}{N}\right)$ & 
${\cal S}^{Q}_{N}(\eta_{\rm opt})$ \\
& & &   \\ 
\hline 
 3      & 1 & 1.5 & 1   \\ 
4 &  2 &  2.83 &  1.85   \\ 
5&  3 & 4.05  & 2.62   \\ 
6 & 4 &  5.20  & 3.35   \\ 
10 & 8  &  9.51  &  6.08   \\ 
20 & 18 & 19.75  &  12.59   \\
50 & 48 &  49.90  &  31.77   \\ 
100 & 98 & 99.95 &  63.62  \\ 
\hline 
\end{tabular}
\caption{ Maximum attainable value ${\cal S}^{Q}_{N}({\eta_{\rm opt}})=\eta_{\rm opt}\, N\, \cos\left(\frac{\pi}{N}\right)$, of the left hand side of  the   $N$ term  temporal correlation inequality (\ref{qtb}),  when the qubit POVMs employed are jointly measurable (see (\ref{achbound})).} 
\end{table}
 
\section{Linear steering inequality and joint measurability} 
    
Quantum steering  (introduced by Schr{\" o}dinger in 1935~\cite{Schrodinger}) has gained much impetus in recent years. In 1989, Reid~\cite{Reid}  proposed an experimentally testable steering critera, which revealed that -- apart from Bell-type non-locality -- steering  is yet another distinct  manifestation of Einstein-Podolsky-Rosen (EPR) non-locality in spatially separated composite quantum systems. A conceptually clear formalism of  EPR steering (in terms of local hidden state (LHS) model) has been formulated by  Wiseman {\em et. al.}~\cite{Wiseman}. They elucidated that steering constitutes a different kind of non-locality, which  lies between entanglement and Bell-type non-locality. Several steering inequalities -- suitable for the experimental demonstration of this form of EPR {\em spooky action at a distance} -- have been derived in Ref.~\cite{Cavalcanti}. Moreover it has been identified that putting steering phenomena to experimental test is much easier compared to  demonstrations of Bell-type non-locality~\cite{Wiseman, Saunders}. Interestingly, steering  framework  is  useful to  investigate the joint measurability problem and vice versa~\cite{Brunner14,Guhne14,Guhne15,HSKJOSAB15,Pusey15,Heinosari15,HSKCurrSc}. In this section, we discuss a linear steering inequality   --  derived by Jones and Wiseman~\cite{Jones} -- where measurements of equatorial qubit observables are employed. We show that this steering inequality exhibits a striking equivalence with the joint measurability of the equatorial  qubit observables, discussed in Sec.~II.

Suppose Alice prepares a bipartite quantum state $\rho_{AB}$ and sends a subsystem to Bob. If the state is entangled, and Alice chooses suitable  local  measurements, on her part of the state, she can affect  Bob's quantum state remotely. How would Bob convince himself that his state is indeed steered by Alice's local measurements?   In order to verify that his (conditional) states  are steered,  Bob asks Alice to perform local measurements  of the observables $\mathbf{X}_k=\sum_{a_k}\, a_k\ \Pi_{x_k}(a_k),\ $ on her part of the state and communicate the outcomes $a_k$ in each experimental trial. If Bob's conditional reduced states (unnormalized)   $\varrho^B_{a_k\vert x_k}={\rm Tr}_A[
\Pi_{x_k}(a_k)\otimes \mathbbm{1}_B\, \rho_{AB}]$,  admit a LHS decomposition~\cite{Wiseman} viz., 
$\varrho_{a_k\vert x_k}=\sum_{\lambda}\, p(\lambda)\,  p(a_k\vert x_k, \lambda)\, \rho^B_\lambda$,  (where $0\leq p(\lambda)\leq 1;\ \sum_\lambda p(\lambda)=1$ and $0\leq p(a_k\vert x_k, \lambda)\leq 1; \sum_{a_k}\, p(a_k\vert x_k, \lambda)=1$; $(p_\lambda, \rho^B_\lambda)$ denote Bob's LHS ensemble), then Bob can declare that Alice is not able to steer his state through local measurements at her end. In addition to entanglement being a necessary ingredient (but not sufficient~\cite{Wiseman}), incompatibility of Alice's local measurements too plays a crucial role to reveal steering~\cite{Brunner14, Guhne14, Guhne15}. In the following subsection  we unfold the intrinsic link between steering  and measurement compatibility in a specific two qubit protocol.   

\subsection{Linear steering inequality for a two qubit system}

Consider a qubit observable~\cite{Jones} 
\begin{equation} 
\label{spl}
{\mathbf S}_{\rm plane}=\frac{1}{\pi}\int_{0}^{\pi}\, d\theta\, 
\alpha_\theta\, \sigma_\theta,
\end{equation}
where  $\sigma_\theta=\sigma_{x}\, \cos (\theta) + \sigma_y\, 
\sin (\theta)$ denotes an equatorial qubit observable and $-1\leq \alpha_\theta\leq 1$. Expectation value of the observable  ${\mathbf S}_{\rm plane}$ is upper bounded by  
\begin{eqnarray} 
\label{splineq}
\langle {\mathbf S}_{\rm plane} \rangle &\leq& \frac{1}{\pi}\, 
\int_{0}^{\pi}\, d\theta\, \langle \sigma_\theta\rangle \nonumber \\ 
&=&\frac{1}{\pi}\, 
\int_{0}^{\pi}\, d\theta\,
 \left( \langle \sigma_x\rangle \cos (\theta)+
\langle \sigma_y\rangle\sin (\theta) \right)\nonumber \\ 
 &=& \frac{2}{\pi} \langle \sigma_y\rangle \nonumber \\
\Rightarrow && \langle{\mathbf S}_{\rm plane}\rangle \leq \frac{2}{\pi} 
\end{eqnarray}

Suppose Alice and Bob share a two qubit state $\rho_{AB}$; Bob asks Alice to perform measurements of $\sigma^{A}_{\theta}$ and communicate the outcome $a_{\theta}=\pm 1$ of her measurements. After Alice's measurements,  Bob will be left with an ensemble $\{p(a_{\theta}\vert\theta), \rho^{B}_{a_{\theta}}\}$ where 
$\rho^{B}_{a_{\theta}\vert \theta}={\rm Tr}_A[\Pi_{\theta}
(a_{\theta})\otimes \mathbbm{1}_B\, \rho_{AB}]/p(a_{\theta}\vert\theta);$ 
$p(a_{\theta}\vert\theta)={\rm Tr}[\Pi_{\theta}(a_{\theta})\otimes 
\mathbbm{1}_B\rho_{AB}]$  denote Bob's conditional states (here $\{\Pi_\theta(a_\theta=\pm 1)\}$ denote PV measurements of the observable $\sigma_\theta$).    At his end, Bob would then  measure the observable $\sigma^{B}_{\theta}$.
Suppose he gets an outcomes $b_{\theta}=\pm 1$ with probability 
$p(b_{\theta}\vert a_{\theta};\theta)=
{\rm Tr}[\Pi_{\theta}(b_{\theta})\rho^{B}_{a_{\theta}
\vert \theta}]$. He evaluates the {\em conditional expectation value} of the observable $\sigma^{B}_{\theta}$ -- based on the statistical data he obtains -- as follows:  
\begin{equation}
\label{cav} 
\langle \sigma^{B}_{\theta}\rangle_{a_{\theta}}= 
\sum_{b_{\theta}=\pm 1}\, b_\theta\,  p(b_{\theta}\vert 
 a_{\theta};\theta). 
\end{equation} 
If the conditional probabilities $p(b_{\theta}\vert  
a_{\theta};\theta)$ originate from  a LHS model i.e., if 
\begin{eqnarray}
p(b_{\theta}\vert  a_{\theta};\theta)&=&
\sum_{\lambda}\, p(\lambda)\,  p(a_{\theta}\vert \theta,  
\lambda)\, {\rm Tr}[\Pi_{\theta_B}(b_{\theta})\rho^B_\lambda] \nonumber \\ 
&=&\sum_{\lambda}\, p(\lambda)\,  p(a_{\theta}\vert \theta, \lambda)\, \langle \Pi_{\theta}(b_{\theta})\rangle_\lambda, 
\end{eqnarray}
(where we have denoted $\sum_{b_{\theta}=\pm 1} \, b_\theta\,
\langle \Pi_{\theta}(b_{\theta})\rangle_\lambda=\langle 
\sigma^B_{\theta}\rangle_\lambda$), one gets the conditional expectation value in the LHS model as follows: 
\begin{eqnarray}
\langle \sigma^B_{\theta}\rangle_{a_{\theta}
\vert \theta} &=& 
\sum_{\lambda}\, p(\lambda)\,  
p(a_{\theta}\vert\theta, \lambda)\, 
\left\{\sum_{b_{\theta}=\pm 1} \, b_\theta\,\langle \Pi_{\theta}(b_{\theta})
\rangle_\lambda\right\}\nonumber \\
&=& \sum_{\lambda}\, p(\lambda)\,  p(a_{\theta}\vert\theta, \lambda)\, 
\langle \sigma_{\theta}\rangle_\lambda.
\end{eqnarray}

Whenever LHS model holds, the  inequality 
\begin{equation}
\label{lsi}
 \frac{1}{\pi}\, 
\int_{0}^{\pi}\, d\theta\, \alpha_\theta\, \langle \sigma^B_
{\theta}\rangle_{a_{\theta}\vert \theta} \leq \frac{2}{\pi} 
\end{equation} 
is obeyed,  for any $-1\leq \alpha_\theta\leq 1$ in the LHS framework. 
Now, denoting
$\sum_{a_{\theta}=\pm 1}\, a_{\theta}\, p(a_{\theta}\vert\theta)\, 
 \langle \sigma^B_{\theta} \rangle_{a_{\theta}
\vert \theta} = 
\langle \sigma^A_{\theta}\, \sigma^B_{\theta}\, \rangle$  one obtaines the linear steering inequality~\cite{Jones}
\begin{equation}
\label{plstin}
 \frac{1}{\pi}\, 
\int_{0}^{\pi}\, d\theta\, \langle \sigma^{A}_{\theta}\,  
\sigma^{B}_{\theta}\rangle \leq \frac{2}{\pi}. 
\end{equation} 
Violation of the inequality (\ref{plstin}) in any bipartite quantum state $\rho_{AB}$ demonstrates non-local EPR steering phenomena (more specifically, violation implies  falsification of the LHS model, which  confirms that Alice can indeed steer Bob's state remotely via her local measurements). 

 Note that implementing  infinite number of measurements (i.e., measurement of 
$\sigma^B_{\theta}$ by Bob conditioned by  the outcomes of  Alice's measurement of $\sigma^A_{\theta}$, in the entire equatorial half plane $0\leq \theta\leq \pi$), is a tough  task in a realistic experimental scenario. So, it would be suitable to consider a finite-setting of  $N$ evenly spaced equatorial measurements  of 
$\sigma_{\theta_{k}}$ \ \ \ \   (such that the successive angular separation is given by $\pi/N$ i.e.,  
$\theta_{k+1}-\theta_k=\pi/N$) by Bob,  conditioned by the $\pm 1$ valued outcomes $a_k$ of  Alice's measurements $\sigma^A_{\theta_k}$. This leads to the following linear steering inequality in the finite setting~\cite{Jones}: 
\begin{equation}
\label{finiteN}
\frac{1}{N}\sum_{k=1}^{N}  \langle \sigma^A_{\theta_k}\,
\sigma^B_{\theta_{k}}\rangle \leq f(N)
\end{equation} 
where  
\begin{equation}
\label{fn}
f(N)=\frac{1}{N}\left(\left\vert \sin\left(\frac{N\pi}{2}\right) \right\vert+ 2\sum_{k=1}^{[N/2]}\, \sin\left[(2k-1)\frac{\pi}{2N}\right]\right), 
\end{equation}
corresponds to the maximum eigenvalue of the observable 
$\frac{1}{N}\sum_{k=1}^{N}\, \sigma_{\theta_{k}}$. 

One obtains  $f(2)=1/\sqrt{2}$, 
$f(3)\approx 0.6666$, $f(4)=0.6533$, $f(10)\approx 0.6392$ for  smaller values of $N$. 
(Note that there is a striking match between the degree of incompatibility 
$\eta_{\rm opt}$ listed in Table.~II and the upper bound $f(N)$ of the inequality (\ref{finiteN})).  The factor $f(N)\rightarrow 2/\pi\approx 0.6366$  in the limit $N\rightarrow\infty$.  

We discuss the violation of the steering inequality (\ref{finiteN}) when Alice and Bob share a maximally entangled two qubit state. 

\subsection{Violation of the linear steering inequality by a two qubit  maximally entangled state}

Let Alice and Bob share a maximally entangled Bell state 
$\vert \psi^-\rangle=(1/\sqrt{2})\, [\vert 0_A,\, 1_B\rangle
-\vert 1_A,\, 0_B\rangle]$. Alice performs PV measurement $\{\Pi_{\theta_k}(a_k)=\frac{1}{2}\, (\mathbbm{1}+ a_k\, \sigma_{\theta_k})\}$ of one of the equatorial qubit observable $\sigma_{\theta_k}$,   which results in an outcome $a_k=\pm 1$, leaving     
Bob's conditional state in the form:   
\begin{eqnarray}
\label{cond}             
\rho^B_{a_k\vert\theta_k}&=&{\rm Tr}_A
\left[ \Pi_{\theta_k}(a_k) \otimes \mathbbm{1}_B \vert \psi^-\rangle \langle \psi^-\vert\right]/p(a_k\vert\theta_k) \nonumber \\ 
&=&  \Pi_{\theta_k}(a_k).
\end{eqnarray} 
( Alice's outcomes $a_k=\pm 1$ are totally random and occur with probability $p(a_k\vert\theta_k)=1/2$ for any measurement setting $\theta_k$).

Bob then performs sharp measurements $\{\Pi_{\theta_k}(b_k)\}$ on his   state and computes the conditional average value of the observable 
$\sigma^{B}_{\theta_k}$ to obtain, 
\begin{eqnarray}
\langle \sigma^{B}_{\theta_k}\rangle_{a_k\vert\theta_k}&=& \sum_{b_k=\pm 1} \, 
b_k\, {\rm Tr}[\Pi_{\theta_k}(a_k)\, \Pi_{\theta_k}(b_k)]\nonumber \\
&=& a_k 
\end{eqnarray}    
Further, evaluating the  average of  $\langle \sigma^{B}_{\theta_k}\rangle_{a_k\vert\theta_k}$ together with Alice's outcomes $a_k$, one obtains, 
\begin{eqnarray} 
\langle \sigma^A_{\theta_k}\,
\sigma^B_{\theta_{k}}\rangle&=&\sum_{a_k=\pm 1}\, a_k\, 
p(a_k)\, \langle \sigma^{B}_{\theta_k}\rangle_{a_k\vert\theta_k}\nonumber \\ 
&=& 1    
\end{eqnarray} 
Thus, the left hand side of the linear steering inequality (\ref{finiteN}) may be readily evaluated and it is given by \break  
$\frac{1}{N}\sum_{k=1}^{N}\langle \sigma^A_{\theta_k}\,\sigma^B_{\theta_{k}}\rangle=1$, which is clearly larger than the upper bound $f(N)$ of the steering inequality (note that   
$f(N)$   varies from its largest   $f(2)\approx 0.7071$ for $N=2$ measurement settings to its limiting value   $f(\infty)=0.6366$ when $N\rightarrow\infty$). 
In the next subsection we show that the violation of the  steering inequality reduces to an inequality $\eta>\eta_{\rm opt}$ (i.e., the unsharpness parameter $\eta$ of Alice's local equatorial qubit POVMs exceeds the cut-off value $\eta_{\rm opt}$ specifying  their compatibility) which, in turn, implies that the set of Alice's measurements are incompatible.    

It is pertinent to point out that a modification of the finite setting  linear steering inequality (\ref{finiteN}) --  violation of which has been tested experimentally~\cite{WisemanNature}: Including a single nonequatorial  measurement of $\sigma_z$ by Bob, the linear steering inequality (\ref{finiteN})-- constructed for a finite set of equatorial observables -- gets  modified into a non-linear steering inequality~\cite{Jones},  violation of which is shown to be more feasible for experimental detection, than that of its linear counterpart)~\cite{Jones}. In an ingenious experimental set up~\cite{WisemanNature},  where a single photon is split into two ports by a beam-splitter, it has been rigorously demonstrated that a set of six different equatorial measurements in one  port (i.e., Alice's end) can indeed steer the state of the photon in the other port (Bob's end).     

\subsection{Joint measurability condition from linear steering inequality} 

Now, we proceed to discuss the implications of joint measurability on the linear steering inequality (\ref{finiteN}). 

If Alice performs unsharp measurement of one of the equatorial qubit POVMs 
$\mathbbm{E}_{\theta_k}=
\{E_{\theta_k}(a_k)=\frac{1}{2}\, (\mathbbm{1}+\eta\, a_k\, \sigma_{\theta_k})\}$   with an outcome $a_k=\pm 1$,  
Bob is left with the following  conditional state,    
\begin{eqnarray}
\label{unsharpcond}             
\rho^B_{a_k\vert\theta_k}&=&{\rm Tr}_A
\left[ (E_{\theta_k}(a_k) \otimes \mathbbm{1}_B)\, \vert \psi^-\rangle \langle \psi^-\vert\right]/p(a_k\vert\theta_k) \nonumber \\ 
 &=& \frac{1}{4} \left(\mathbbm{1}-\eta\, a_k\, \sigma_{\theta_k}\right)
/p(a_k\vert \theta_k)\nonumber \\
&=& \bar{E}_{\theta_k}(a_k),
\end{eqnarray} 
the probability of Alice's obtaining the outcome $a_k$ being  $p(a_k\vert \theta_k)=1/2$. Here, we have denoted the {\it spin-flipped} version of the POVM  $\{E_{\theta_k}(a_k)=(\mathbbm{1}+\eta\, a_k\, \sigma_{\theta_k})/2\}$ by 
$\{\bar{E}_{\theta_k}(a_k)=(\mathbbm{1}-\eta\, a_k\, \sigma_{\theta_k})/2\}.$
Following Alice's measurement, Bob carries out sharp measurements  
$\{\bar{\Pi}_{\theta_k}(b_k)=(\mathbbm{1}-b_k\, \sigma_{\theta_k})/2\}$ on his state and computes the conditional average value of the observable 
$\sigma^{B}_{\theta_k}$ to obtain, 
\begin{eqnarray}
\langle \sigma^{B}_{\theta_k}\rangle_{a_k\vert\theta_k}&=& \sum_{b_k=\pm 1}b_k\,  {\rm Tr}[\rho^B_{a_k\vert\theta_k}\, 
\bar{\Pi}_{\theta_k}(b_k)] \nonumber \\ 
&=& \sum_{b_k=\pm 1}\, b_k\,  {\rm Tr}[\bar{E}_{\theta_k}(a_k)\, 
(b_k)] \nonumber \\
&=& \eta\, a_k.   
\end{eqnarray}    
Averaging the conditional expectation value  $\langle \sigma^{B}_{\theta_k}\rangle_{a_k\vert\theta_k}$  with  Alice's outcomes $a_k$, we obtain,  
\begin{eqnarray} 
\langle \sigma^A_{\theta_k}\,
\sigma^B_{\theta_{k}}\rangle=\sum_{a_k=\pm 1}\, a_k\, 
p(a_k\vert \theta_k)\, \langle \sigma^{B}_{\theta_k}\rangle_{a_k\vert\theta_k}=\eta.    
\end{eqnarray} 
Thus the finite setting linear steering inequality (\ref{finiteN}) reduces to,  
\begin{equation} 
\label{stjoint}
\eta\leq f(N).  
\end{equation}
This  reduces to the  joint measurability condition 
$\eta\leq \eta_{\rm opt}$ for Alice's local unsharp measurements, as one can identify a striking agreement 
between the degree of incompatibility $\eta_{\rm opt}$   (given by (\ref{etaeq}) and listed  in Table.~II)) and the upper bound $f(N)$ (given in  (\ref{fn})) of the finite setting linear steering inequality (\ref{finiteN}).  This is a clear example of the  
 intrinsic connection (established in Refs.~\cite{Brunner14,Guhne14,Guhne15}) between steering  and  measurement incompatibility. Moreover, the equivalence between the degree of incompatibility (as given in (\ref{etaeq}))  and  the linear steering inequality in the finite setting  (see (\ref{stjoint})) highlights the relation between a {\em local} quantum feature i.e., non-joint measurability  and a  {\em non-local} one viz., steerability.  Would it be possible to demonstrate  measurement incompatibility,  without  employing a non-local resource (i.e., an entangled state)? In this direction, it is pertinent to point out that {\em time-like anologues} of steering have been formulated 
recently~\cite{Chen14,HSKJOSAB15}; and there has been an ongoing research interest towards developing resource theories of  measurement incompatibility and non-local steerabililty~\cite{Pusey15,Heinosari15,Chen16}. This leads us to formulate  (in the next subsection) a {\it local  analogue} of the linear steering inequality (\ref{finiteN}) in a single qubit system -- violation of which implies  incompatibility of the qubit POVMs employed in first measurements of the sequential pair.  

\subsection{Local analogue of the linear steering inequality}

As has been discussed in previous subsections, expectation value of  the qubit observale ${\bf S}_{\rm plane}=(1/\pi)\,\int_{0}^{\pi}\, d\theta \alpha_\theta\, \sigma_{\theta};  -1\leq \alpha_\theta\leq 1$,  is bounded by $2/\pi$  (see  (\ref{splineq})). This bound is {\em not} obeyed, in general, if the expectation value of the  observable $\langle\sigma_{\theta}\rangle$ is replaced by its conditional expectation value $\langle \sigma_{\theta}\rangle_{a_\theta\vert \theta}$, evaluated in a sequential measurement, with the first measurement resulting in an outcome $a_\theta$. In particular, in the setting where  finite number of pairwise sequential measurements of the equatorial qubit observable $\sigma_{\theta_k}$, with same angle $\theta_k$, are carried out~\cite{refnote}, the analogue of the steering  inequality (\ref{finiteN}) 
\begin{equation} 
\label{tst}
\frac{1}{N}\, \langle\sigma^{(1)}_{\theta_k}\,\sigma^{(2)}_{\theta_k}\rangle \leq f(N)
\end{equation}
could get violated in the single qubit system.   Here, we have denoted $\langle\sigma^{(1)}_{\theta_k}\,\sigma^{(2)}_{\theta_k}\rangle=\sum_{a_k}\,a_k p(a_k\vert\theta_k)\,\, 
\langle \sigma^{(2)}_{\theta_k}\rangle_{a_k\vert \theta_k}$; \ $\langle 
\sigma^{(2)}_{\theta_k}\rangle_{a_k\vert \theta_k}=\sum_{b_k}\,b_k\,  
p(b_k\vert a_k,\theta_k)$ denotes the conditional expectation value of $\sigma_{\theta_k}$ -- given that the first measurement  has resulted in an outcome $a_k$ with probability $p(a_k\vert \theta_k)$. Now, we proceed to  identify explicitly that violation of the inequality (\ref{tst}) is   merely a  consequence of  measurement incompatibility.            

Consider sequential measurements of equatorial qubit observables $\sigma_{\theta_k}$ in a single qubit state $\rho=\frac{1}{2}\mathbbm{1}$. 
Suppose an unsharp measurement $\{E_{\theta_k}(a_k)=(1/2)\, [\mathbbm{1}+\eta\, a_k\, \sigma_{\theta_k}]\}$ results in an outcome $a_k$, with probability 
$p(a_k\vert\theta_k)={\rm Tr}[\rho\, E_{\theta_k}(a_k)]=1/2$. Correspondingly, the state undergoes a transformation 
\begin{eqnarray}
\rho\rightarrow  \rho_{a_k\vert \theta_k} 
 &=&  E_{\theta_k}(a_k)
\end{eqnarray}
 after the first measurement. Following this with another sharp PV measurement 
$\{\Pi_{\theta_k}(b_k)=(1/2)\, [\mathbbm{1}+b_k\, \sigma_{\theta_k}]\}$, the resulting post measured state takes the form,  
\begin{equation}
\rho_{b_k\vert a_k; \theta_k}= \left[\Pi_{\theta_k}(b_k)\, 
\rho_{a_k\vert\theta_k}\, \Pi_{\theta_k}(b_k)\right]/p(b_k\vert a_k), 
\end{equation}     
where 
\begin{eqnarray}
p(b_k\vert a_k)&=&{\rm Tr}[\rho_{a_k\vert\theta_k}\, \Pi_{\theta_k}(b_k)]\nonumber \\ 
&=& \frac{1}{2}[1+\eta\, a_k\, b_k]
\end{eqnarray}
 is the conditional probability of obtaining the outcome $b_k$ in the second measurement.   
The conditional expectation value of the observable $\sigma_{\theta_k}$ in the second measurement is then evaluated to obtain, 
\begin{eqnarray} 
\langle\sigma^{(2)}_{\theta_k}\rangle_{a_k\vert \theta_k}&=&\sum_{b_k}\,b_k\,  p(b_k\vert a_k,\theta_k)\nonumber \\ 
&=& \eta\, a_k. 
\end{eqnarray}
The average value  $\langle\sigma^{(1)}_{\theta_k}\,\sigma^{(2)}_{\theta_k}\rangle$, evaluated  using the statistical data of the first measurement results in, 
\begin{eqnarray}
\langle\sigma^{(1)}_{\theta_k}\,\sigma^{(2)}_{\theta_k}\rangle&=&\sum_{a_k=\pm 1}\, a_k\, p(a_k\vert \theta_k)\,  
\langle\sigma^{(2)}_{\theta_k}\rangle_{a_k\vert \theta_k} \nonumber \\
&=& \eta.
\end{eqnarray}  
Thus, the inequality (\ref{finiteN}) reduces to $\eta\leq f(N)$, when  $N$ pairs of unsharp-sharp  measurements are carried out sequentially  in a single qubit system. Clearly, the inequality is violated, when only sharp PV measurements (with $\eta=1$) are carried out. On the other hand, the inequality is always obeyed, when  the set  $\{E_{\theta_k}(a_k),\ k=1,2,\ldots N$ of all POVMs, employed in the  first measurements of every sequential pair measurements,  is jointly measurable. In other words, we have shown that violation of the local analogue of the steering inequality (\ref{tst}) in a single qubit system is a consequence of incompatibility of measurements of the qubit POVMs employed in first measurements of the sequential scheme. 

\section{Conclusions}  

Discerning the intrinsic connection between quantum non-locality and measurement incompatibility is significant in that it leads to conceptual clarity in understanding different manifestations of non-classicality. An interesting recent result by Banik {et. al.}~\cite{Banik}, revealed that the degree of measurement incompatibility --  quantifying joint measurability of two dichotomic observables --  places restrictions on  the maximum strength of violation of the CHSH-Bell inequality. A natural question then is  whether such  a quantitative connection exists in general, when more than two measurement settings are involved. In this paper we have  explored the connection between the maximum achievable bound (Tsirelson-like quantum bound) on the violation of $N$ term pairwise correlation inequality~\cite{Guhne13} and the degree of measurement incompatibility of  $N$ dichotomic qubit POVMs, employed in carrying out first measurement of  sequential pair measurements. To this end, we have constructed $N$ term chained correlation inequalities based on the positivity of a sequence of $4\times 4$ moment matrices in the classical probability setting. Replacing the classical dichotomic random variables by qubit observables and classical probability distribution by quantum state, we obtain the analogue of chained $N$ term correlation inequalities in the quantum scenario; in general the correlations do not obey the classical bound -- resulting in the violation of the inequalities. Maximum achievable  quantum bound (Tsirelson-like bound) on one of these chained inequalities  --  involving  pairwise  correlations of statistical outcomes of dichotomic observables measured sequentially in a   single quantum system --  is known~\cite{Guhne13}; and the dichotomic observables, which result in the  maximum quantum violation of the inequality,  correspond to qubit observables, having equal successive angular separation of $\pi/N$ in a plane. We have shown in this work  that the $N$-term chained inequality (\ref{ch3}) is always obeyed, when the set of all POVMs employed in first measurements of every pairwise correlation term, is compatible. However,  measurement incompatibility of equatorial  qubit POVMs serves, in general, as  a necessary condition. For $N>3$, incompatibility is not sufficient to result in violation of  (\ref{ch3}). To be specific,  a tight relation between the degree of incompatibility  and the maximum strength of quantum violation of the correlation inequality  holds mainly in two special cases:  (i) Measurements of a pair of dichotomic observables on one part of a bipartite quantum system are considered. In this case, the degree of incompatibility $\eta_{\rm opt}=1/\sqrt{2}$ -- for  the pair of dichotomic observables to be jointly measurable --  places an upper bound 
$2/\eta_{\rm opt}=2\sqrt{2}$ on the maximum achievable quantum bound of CHSH-Bell inequality~\cite{Banik}) (ii) In a three term  correlation inequality (\ref{ch3}), with a classical upper bound 
1; here, sequential pairwise measurements of $N=3$ dichotomic observables are carried out in a single qubit system prepared initially in a maximally mixed state. The inequality is known to be violated maximally ( quantum upper bound being $3/2$), when the three  dichotomic observables correspond to qubit orientations, forming trine axis (three axes with equal successive angular separations of $\pi/3$ in a plane). In this case, the degree of  measurement incompatibility of the three POVMs is given by $\eta_{\rm opt}=2/3$. When these POVMs are used in first measurements of the sequential pair measurements, the degree of incompatiblity  places restrictions on the maximum achievable quantum bound~\cite{HSKCurrSc} i.e., $1/\eta_{\rm opt}=3/2$. In view of the recent research focus on the equivalence between  joint measurability and non-local steering~\cite{Brunner14,Guhne14,Guhne15},  we have explored  a linear steering inequality -- introduced by Jones and Wiseman~\cite{Jones} -- which involves measurements of $N$ equatorial plane qubit POVMs. We have shown that this indeed unfolds a striking connection between the optimal violation of the $N$ term steering inequality and the degree of incompatibility of equatorial qubit POVMs. 

Within the perspective of our study, it appears natural to ask {\it if one can device a local test (by carrying out a set of sequential measurements on a single quantum system)  to infer about measurement incompatibility -- than employing a non-local steering protocol (which requires an entangled state)?}  We have addressed this question  -- by restricting to  the specific example pertaining to   $N$ equatorial qubit observables -- and have shown that  a local analogue of the linear steering inequality of Ref.~\cite{Jones} can be formulated in a single quantum system -- involving a  linear combination of pairwise conditional correlations, resulting from $N$ sequentially ordered unsharp-sharp pairwise measurements (performed in independent statistical trials for each pair, with the input state for every first measurement being $\rho=\mathbbm{1}/2$) of equatorial qubit observables. Violation of this {\em local} steering inequality is shown to be a reflection of measurement incompatibility of POVMs employed in first of the sequential pairwise measurements.

\section*{Acknowledgements}
We thank the anonymous referee for critical comments, especially on the unsharp-sharp pairwise sequential measurement scheme, which made us take a re-look and revise our manuscript. We also thank helpful discussions with T. Heinosaari. ARU is supported by the University Grants Commission (UGC) Major Research Project (Grant No. MRP-MAJOR-PHYS-2013-29318), Government of India. Local hospitality and facilities offered by Statistical Mathematics Unit, Indian Statistical Institute, Delhi, where this research work  was partly carried out during ARU's visit, is gratefully acknowledged. JPT acknowledges support by UGC-BSR, India.  


\begin{thebibliography}{99}
\bibitem{note1} Joint measurability requires  that in every state, there exists underlying joint probabilities corresponding to the measurement outcomes of a set of  compatible POVMs; and these joint probabilities yield correct marginal probabilities corresponding to the outcomes of all the individual POVMs -- which comprise of a compatible set.
\bibitem{notebell} One can construct  chained $2N$ term  CHSH-Bell correlation inequalities (with  $N$ dichotomic measurement settings each by Alice and Bob) based on positivity of $2N$ sequences of moment matrices -- in an analogous manner in which ((\ref{ch1})-(\ref{ch4})) are derived. The maximum  strength of violation (Tsirelson-like bound) of $2N$ term CHSH inequalities in quantum theory has been derived by Wehner in Ref.~\cite{Wehner}. And the pairwise correlations of oucomes -- registered in local measurements carried out by   Alice and Bob  at their respective ends --  which lead to maximum violation -- arise from sharp PV measurements of   equatorial qubit observables $\sigma^{A,B}_{\theta_k}~=~\sigma^{A, B}_{x}\, \cos(\theta_k)+\sigma^{A,B}_{y}\, \sin(\theta_k),$ 
$\theta_k=k\, \pi/N,\  k=1,2,\ldots, N$. It can be shown that when Alice (or Bob) employs compatible measurements, the Bell-CHSH $2N$ term inequalities do not get violated.  However, measurement  incompatibility is only  necessary, but not sufficient for Bell type tests, involving more than $N=2$ measurement settings~\cite{Brunner14, Guhne14}.
\bibitem{refnote} In order to construct an anologue of the linear steering inequality (\ref{finiteN}) in a single qubit system, sequential measurements of pairs of observables with same angle $\theta_k$ has to be carried out.
\end{thebibliography}
\end{document}